\documentclass[12pt,preprint]{aastex}
\usepackage{amsmath}
\usepackage{booktabs}
\usepackage{threeparttable}
\usepackage{rotating}
\usepackage{multirow}
\usepackage{url}
\shorttitle{$\gamma $-ray and X-ray observation on IceCube neutrino}

\begin{document}

\title{Search for GeV and X-ray flares associated with the IceCube track-like neutrinos }

\author{Fang-Kun Peng\altaffilmark{1,2}, Xiang-Yu Wang\altaffilmark{1,2}}
 \altaffiltext{1}{School of Astronomy and Space Science, Nanjing University, Nanjing 210093, China;
xywang@nju.edu.cn} \altaffiltext{2}{Key laboratory of Modern Astronomy and Astrophysics (Nanjing University), Ministry of Education, Nanjing 210093, China}

\begin{abstract}
Dozens of high-energy neutrinos have been detected by the IceCube
neutrino telescope, but no clear association with any classes of
astrophysical sources has been identified so far. Recently,
\citet{2016arXiv160202012K} report that a PeV cascade-like
neutrino event occurs in positional and temporal coincidence with
a giant gamma-ray flare of the blazar PKS B1424-418. Since IceCube
track-like events have much better angular resolution, we here
search for possible short-term gamma-ray flares that are
associated with the IceCube track-like events with \textsl{Fermi}
Large Area Telescope (LAT) observations.  Among them, three
track-like neutrino events occur within the field of view of
\textsl{Fermi}-LAT at the time of the detection, so search for the
{\em prompt} gamma-ray emission associated with neutrinos are
possible. Assuming a point source origin and a single power law
spectrum for the possible gamma-ray sources associated with
neutrinos, a likelihood analysis of 0.2-100 GeV photons observed
by \textsl{Fermi}-LAT on the timescales of $\sim 12$ hours and one
year are performed, and for the three special neutrinos, the
analysis are also performed on the timescales of thousand of
seconds before and after the neutrino detection. No significant
GeV excesses over the background are found and the upper limit
fluxes at 95\% confidence level are obtained for different
timescales. We also search for possible  hard
X-ray transient sources associated with the IceCube track-like
neutrino events, but the search also yields null results.
{We discuss the implication of the non-detection of
gamma-ray flares for the constraints on the neutrino source
density.}
\end{abstract}

\keywords{neutrinos--galaxies: active--gamma rays}

\section{Introduction}
The IceCube telescope has detected TeV-PeV neutrinos from
extraterrestrial sources  for the first time
\citep{2013Sci...342E...1I,2014PhRvL.113j1101A,2015arXiv151005223T},
which open a new window to explore the high-energy universe. The
explanation of a single atmospheric origin of these high-energy
neutrino events collected during 4 years has been strongly
disfavored at around $6.5 \sigma $ level of confidence \citep{2014PhRvL.113j1101A,2015arXiv151005223T} .
The best fit result for the high-energy astrophysical
neutrino flux reaches a level of $E_{\nu}^2\Phi_{\nu}\sim 10^{-8}$
GeV cm$^{-2}$ s$^{-1}$ sr$^{-1}$ per flavor between around 60 TeV
and 2 PeV, and the spectral index of the power law model is $-2.5
\sim -2.0 $ \citep{2014PhRvL.113j1101A,2015ApJ...809...98A}.

The astrophysical neutrinos have shown no significantly
directional clustering \citep{2014PhRvL.113j1101A}. They also show
no clear association with any known classes of astrophysical
sources so far. High-energy neutrino emission results from the
decays of charged pions produced in the interaction between
relativistic protons and ambient gas ($pp$) or ambient radiation
($p \gamma $), and the same processes inevitably  produce
high-energy gamma-ray photons via the neutral pion decays.
Although very high-energy photons above, e.g. 100 GeV, may be
absorbed in the source or during the propagation in the
intergalactic space, GeV photons could escape and  arrive at the
Earth accompanying neutrinos.  The potential astrophysical sources
to produce high-energy neutrinos and photons include star-forming
galaxies
\citep{2006JCAP...05..003L,2013PhRvD..87f3011H,2013PhRvD..88l1301M,2014PhRvD..89l7304A,2014PhRvD..89h3004L,2014JCAP...09..043T,2014JCAP...11..028W,2015ApJ...805...95C},
tidal disruption events \citep{2016PhRvD..93h3005W}, gamma-ray
bursts
(GRBs)\citep{1997PhRvL..78.2292W,2013JCAP...06..030C,2013ApJ...766...73L,2013PhRvL.111l1102M},
active galactic nucleus
(AGNs)\citep{2008APh....29....1A,2013PhRvL.111d1103K,2013PhRvD..88d7301S,2014PhRvD..90b3007M,2014JHEAp...3...29D},
 double white dwarf mergers \citep{2016arXiv160808150X} and even
Galactic sources \citep{2013ApJ...774...74F,2013PhRvD..88h1302R,2014PhRvD..90b3010A,2014PhRvD..90b3016L,2014PhRvD..89j3002N}, see
\citet{2015RPPh...78l6901A} for a review. However, combined data
analysis between IceCube neutrinos events and $\gamma $-ray
sources sample, such as GRBs, AGNs, soft $\gamma $-ray repeaters,
supernova remnants, pulsars, microquasars, and X-ray binaries
\citep{2014MNRAS.443..474P,2014A&A...566L...7K,
2015ApJ...809...98A,2015ApJ...807...46A,2016MNRAS.457.3582P,2016EPJWC.12105006G,2016ApJ...824..115A,2016SCPMA..59a5759W},
do not reveal any firm associations up to now.

Recently \citet{2016arXiv160202012K} find that a cascade-like PeV
neutrino event  occurs in positional and temporal coincidence with
a giant gamma-ray flare of the flat spectrum radio quasars (FSRQs)
PKS B1424-418, with a chance probability of $5\%$ for such
coincidence (i.e., a 2$\sigma$ confidence level correlation). This
cascade-like neutrino has an angular error of $\sim 15^{\circ}$.
In contrast to cascade-like events, the median angular resolution
of muon track neutrino events are much better ($\simeq1^{\circ}$),
and hence they are good candidates to search for the
electromagnetic counterparts. \citet{2015MNRAS.451..323B} have
performed the search for the gamma-ray counterparts of the first 7
IceCube track-like neutrinos ($E_{\nu} > 30$ TeV) using 70-month
\textsl{Fermi}-LAT data, and no steady $\gamma$-ray counterparts
are found. For the purpose of examining whether short-term
transient sources like PKS B1424-418 are associated with
neutrinos, we here search for possible transient gamma-ray
counterparts (on the timescale as short as hours) of 12 track-like
events\footnote{\url{http://gcn.gsfc.nasa.gov/amon_hese_events.html,http://gcn.gsfc.nasa.gov/amon_ehe_events.html}}
observed by IceCube up to August 6, 2016
\citep{2014PhRvL.113j1101A,2015arXiv151005223T,2015ATel.7856....1S,2016GCN..19363...1B}.
We select neutrino events with energies larger than 60 TeV to
reduce the contamination from the  atmospheric  background
\citep{2014PhRvL.113j1101A}. This analysis method is rather than
cross-correlating with known catalogs, but uses the
\textsl{Fermi}-LAT survey data to search for new $\gamma $-ray
transients related with the IceCube track-like events or flux
variability of known $\gamma $-ray sources. Different from
\citet{2015MNRAS.451..323B},  our work aims to find possible
short-term or prompt GeV emission  associated with these
neutrinos, such as X-ray transients or gamma-ray transients (e.g.
AGN gamma-ray flares), which could be missed in the 70-month long
timescale analysis similar to \citet{2015MNRAS.451..323B}. {We
note that the gamma-ray flare of PKS B1424-418 is extremely strong
and lasts for more than one year. Sources with flares at such flux
level and durations from neutrino directions examined in
\citet{2015MNRAS.451..323B} would have been observed as a
significant excesses. However, for sources like short-term bright
flare of AGNs or GRBs, the signal may be diluted and become
undetectable in a longer time interval. Our analysis is most
sensitive to gamma-ray sources that are transient only in a short
time interval and are quiescent over a  long time interval. For
example, a radio-intermediate quasars III Zw 2 exhibits distinct
GeV flares in the short term, but no significant gamma-ray signal
has been detected in the time-averaged 7-year \textsl{Fermi}-LAT
data \citep{2016ApJS..226...17L}. \textsl{Fermi} All-sky
Variability Analysis has found that, among 518 flaring gamma-ray
sources, 77 sources lack of gamma-ray counterparts in the 7.4
years of \textsl{Fermi} observations \citep{2016arXiv161203165A}.
Some tidal disruption events (TDEs) show short-duration luminous
X-ray flares. Although high-energy emissions from TDEs have not
been detected by \textsl{Fermi}-LAT so far
\citep{2016ApJ...825...47P}, they are potential sources of giant
gamma-ray flares \citep{2009ApJ...693..329F}. GRBs also have
bright GeV emission during the prompt and afterglow phase,
although there is no evidence of association with IceCube
high-energy neutrinos so far. Our analysis would be sensitive to
such short-term gamma-ray transients. }

The organization of the paper is as follows. In section 2, we
describe the  search results with the \textsl{Fermi}-LAT
observations, the result for the search of the \textsl{Swift} hard
X-ray transient sources is described in section 3. In section 4,
we discuss the implication of the non-detection of gamma-ray
transients for constraining the source number density. Finally we
give the conclusions and discussions in section 5.

\section{\textsl{Fermi} data reduction}
\subsection{\textsl{Fermi}-LAT data analysis}
The newly released \textsl{Fermi}-LAT \citep{2009ApJ...697.1071A}
Pass 8 SOURCE data (P8R2 Version 6) and \textsl{Fermi}  science
tools version v10r0p5 are used in the present work. An unbinned
maximum likelihood analysis is performed on a region of interest
(ROI) with a radius $10^{\circ}$ centered on the right ascension
and declination of the each IceCube track-like neutrino. All
FRONT+BACK converting photons with energies between 0.2-100 GeV
are taken into consideration. We apply the maximum zenith-angle
cut $zmax = 90^{\circ}$ to eliminate the Earth's limb emission.
The expression of (DATA\_QUAL $> 0$) \& (LAT\_CONFIG ==1) is used to further filter the data.
A source model is generated containing the position and spectral
definition for all the point sources and diffuse emission from the
3FGL \citep{2015ApJS..218...23A} within $15^{\circ}$ of the ROI
center. The Galactic and extragalactic diffuse models are
gll\_iem\_v06.fits and iso\_P8R2\_SOURCE\_V6\_v06.txt,
respectively. We add a point source with power-law spectrum
($dN/dE=A\times (E/E_0)^{-\Gamma}$) on each track-like neutrino
position in the source model file. Since we pay attention to the
short-term behavior of $\gamma $-ray emission on the timescales of
hours or months, the spectral indices of all point sources in the
source model file are fixed to their 3FGL catalogue values to
solve convergence problems. The normalization factors of point
sources, the extragalactic diffuse emission, and the Galactic
diffuse emission are left free to vary. After each successful fit,
test-statistic (TS) map centered on the neutrino position is created to check if there is
any  excess $\gamma $-ray emission above the background beyond the
3FGL catalog. All the upper limit fluxes are reported at the $95\%$
confidence level with fixed spectral index $\Gamma =2$.
We have tested that assuming different spectral
indices would result in a slight but insignificant difference.

\subsection{\textsl{Fermi}-LAT data search results}
\begin{figure}
\centering
\includegraphics[scale=0.7]{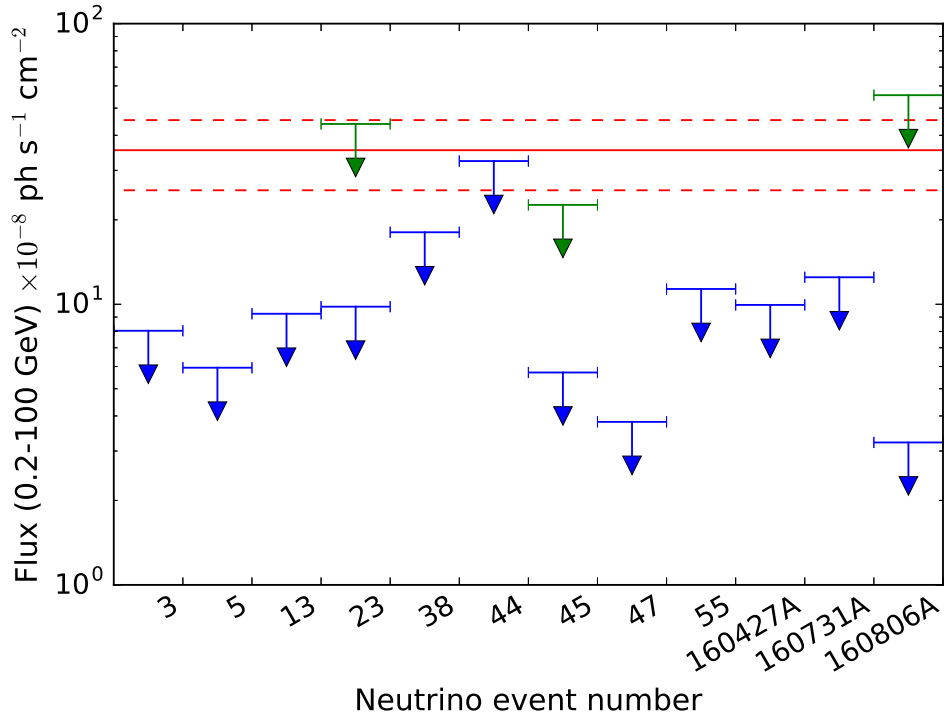}
\caption{Comparison of the upper limit gamma-ray fluxes on the
timescales of 12 hours and $\sim 1000$ s, as reported in Table
\ref{lcdata} (blue data) and Table \ref{lcondata} (green data)
respectively, with the flux of the gamma-ray flare (red lines)
from PKS B1424-418. The dashed red lines indicate the one $\sigma$ flux range
of the gamma-ray flare from PKS B1424-418.} \label{lc8Tfig}
\end{figure}

We first perform the data analysis over $\sim 12 $ hours, i.e. 6
hours before and 6 hours after the neutrino detection time, to
search for possibly prompt GeV emission accompanying these
neutrino events. No significant gamma-ray emissions at the
position of the track-like neutrino events are found, and thus
their upper limit fluxes are obtained, see Table \ref{lcdata} and
Figure \ref{lc8Tfig}. There is no new $\gamma $-ray source around
the region of the neutrino position identified by checking the TS
map. For comparison,  the \textsl{Fermi}-LAT data analysis for the
gamma-ray flaring blazar PKS B1424-418 in a similar time period
{(centering at the detection time of the neutrino
event number 35 )} is also carried out and the result is presented in Table
\ref{lcdata} and Figure \ref{lc8Tfig}. For the same  period of
time, PKS B1424-418 shows bright emission with detection
significance of $TS =57$, and the photon flux is $3.54\pm 0.99
\times 10^{-7}$ ph cm$^{-2}$ s$^{-1}$ (i.e. the corresponding
energy flux is $4.05\pm 1.14 \times 10^{-10}$ erg cm$^{-2}$
s$^{-1}$ in $0.2-100$ GeV). As we can see in Figure \ref{lc8Tfig},
the upper limit fluxes of any possible point sources associated
with the track-like neutrinos are below the flux of the PKS B1424-418.
We also find that all the 3FGL sources within the
 $2R_{50}$ angular error of track-like neutrinos are too weak to  be detected by \textsl{Fermi}-LAT
for twelve-hour observations (here $R_{50} $ means angular error
at the 50\% confidence level).

\begin{figure}
\centering
\includegraphics[scale=0.7]{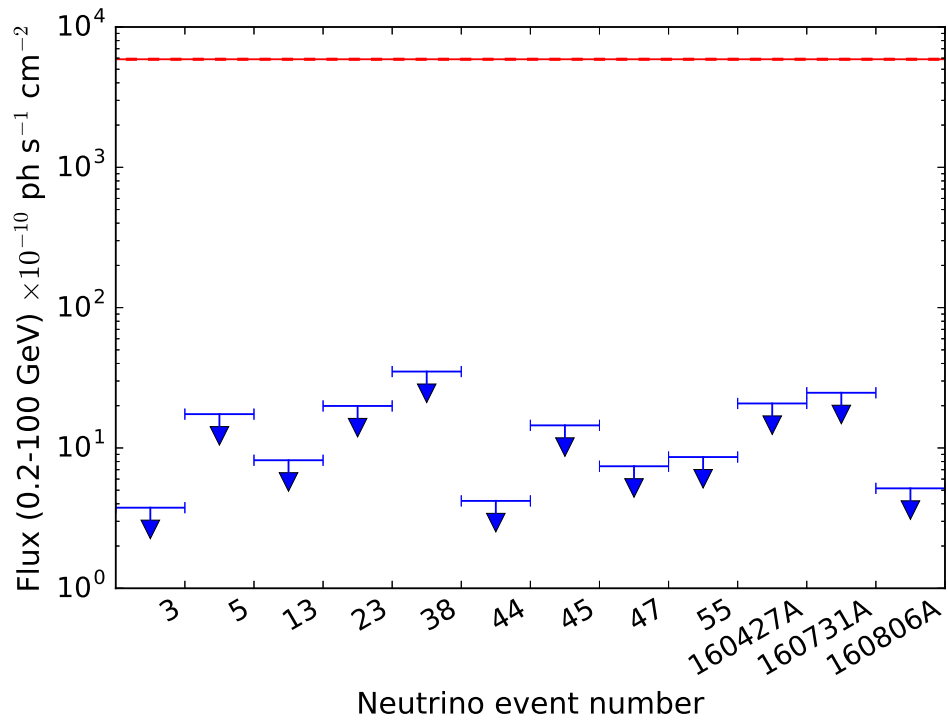}
\caption{Upper limit gamma-ray fluxes of the track-like neutrino
events for one year \textsl{Fermi}-LAT observations, as reported
in Table \ref{lcdata}. The flux of the gamma-ray flare of PKS
B1424-418 (red lines) is presented for comparison.}
\label{lc1yrfig}
\end{figure}

As some blazars outbursts occur on the timescale of months, we
further choose one year for the time window to search for
gamma-ray flares. The likelihood analysis of \textsl{Fermi}-LAT
data of each track-like neutrino event is conducted, which also
yields a null result. The upper limit fluxes covering the period
of half year before and half year after the neutrino detection
time are given in Table \ref{lcdata} and also shown in Figure
\ref{lc1yrfig}. {Similarly, over one year
\textsl{Fermi}-LAT observation centering at the detection time of
the neutrino event number 35, PKS B1424-418 shows high photon flux
$5.89 \pm 0.06 \times 10^{-7}$ ph cm$^{-2}$ s$^{-1}$ ($TS =
50770$), which corresponds to an energy flux of $8.17\pm 0.14
\times 10^{-10}$ erg cm$^{-2}$ s$^{-1}$.} The upper limit
gamma-ray fluxes for the track-like neutrinos are far below the
flux of PKS B1424-418. We note that neutrino event number 5 has a
known 3FGL $\gamma $-ray source J0725.8-0054 (BL Lac object PKS
0723-008) located $\sim 1^{\circ}$ from neutrino's position.
Another 3FGL $\gamma $-ray source J2227.8+0040 (BL Lac object PMN
J2227+0037) is located $\sim 0.7^{\circ}$ from the number 44
neutrino event. The two sources are detected at $TS =83$ and
$TS=35$ in one year observation respectively. The one year fluxes
of the two sources are consistent with the values  published in
the 3FGL catalogue, which are two orders of magnitude lower than
that of PKS B1424-418. While these two 3FGL sources appear to have
a hard spectral index, there is no evidence that they are  TeV
gamma-ray sources (\url{http://tevcat.uchicago.edu/}). Moreover,
when a spatial error of $2R_{50}$ is considered, additional 9 3FGL
sources, most of which are blazars,  are in positional agreement
with the these track-like neutrino events (see Table \ref{FGLdata}
for more details).  Their gamma-ray fluxes  are, however, too low
to account for the observed neutrino flux, in contrast to PKS
B1424-418. No new $\gamma $-ray source around the region of the
neutrino position is discovered.

We calculate the significance of the spatial coincidence of 3FGL
sources with the IceCube track-like neutrinos, running 10000
simulations in which the declination and the right ascension of
each 3FGL sample are randomized. For each simulation, we obtain a
count number $n$ of 3FGL sources within $R_{50}$ of our track-like
neutrino events sample. The chance probability  is calculated as
the ratio between the number of the simulations that have $n
\geqslant 2 $ and the total number of simulations. This approach
results in a chance probability  $\sim 98\%$, suggests that the
coincidence between two 3FGL sources, J0725.8-0054 (PKS 0723-008)
and J2227.8+0040 (PMN J2227+0037), and the track-like neutrino events
is merely by chance. If the declination is fixed and the right
ascension is randomized only, the chance probability of such
spatial coincidence reaches $\sim 83\%$. We therefore find no
evidence of gamma-ray emission associated with the IceCube track-like
neutrino events. Considering the $\gamma$-ray flux limits for one
year \textsl{Fermi}-LAT  observation, we suggest that any
gamma-ray flares that are associated with the IceCube track-like
neutrino events must be at least one order of magnitude dimmer
than that of PKS B1424-418\footnote{Interestingly,
\citet{2016arXiv161005306G} find that a hybrid model with
sub-dominant hadronic component is needed to explain the
multi-waveband observation of PKS B1424-418 flare.} (see Figure
\ref{lc1yrfig}).

\subsection{IceCube neutrino events number 23, 45 and 160806A }
\begin{figure}
\centering
\includegraphics[angle=0,scale=0.3]{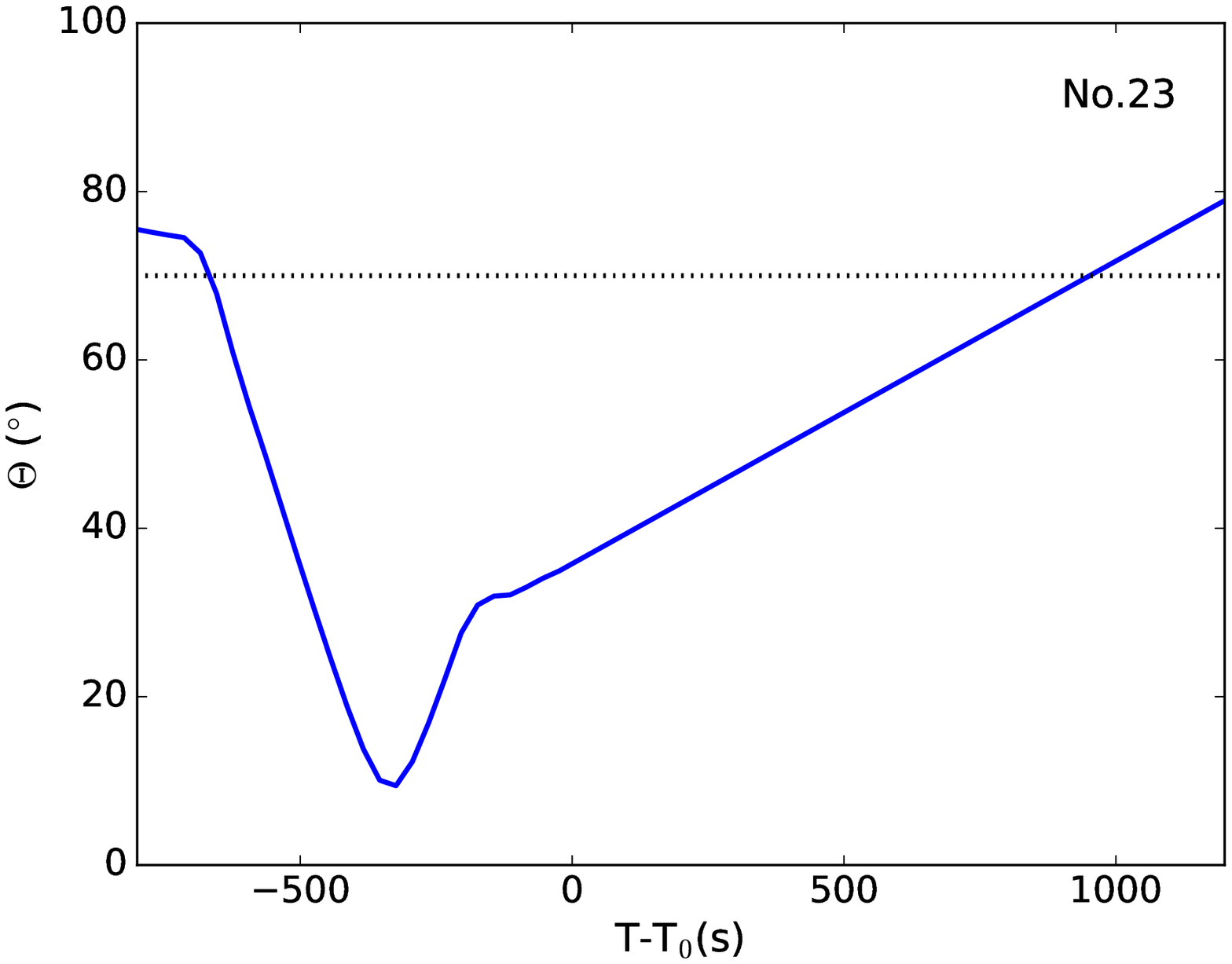}%
\includegraphics[angle=0,scale=0.3]{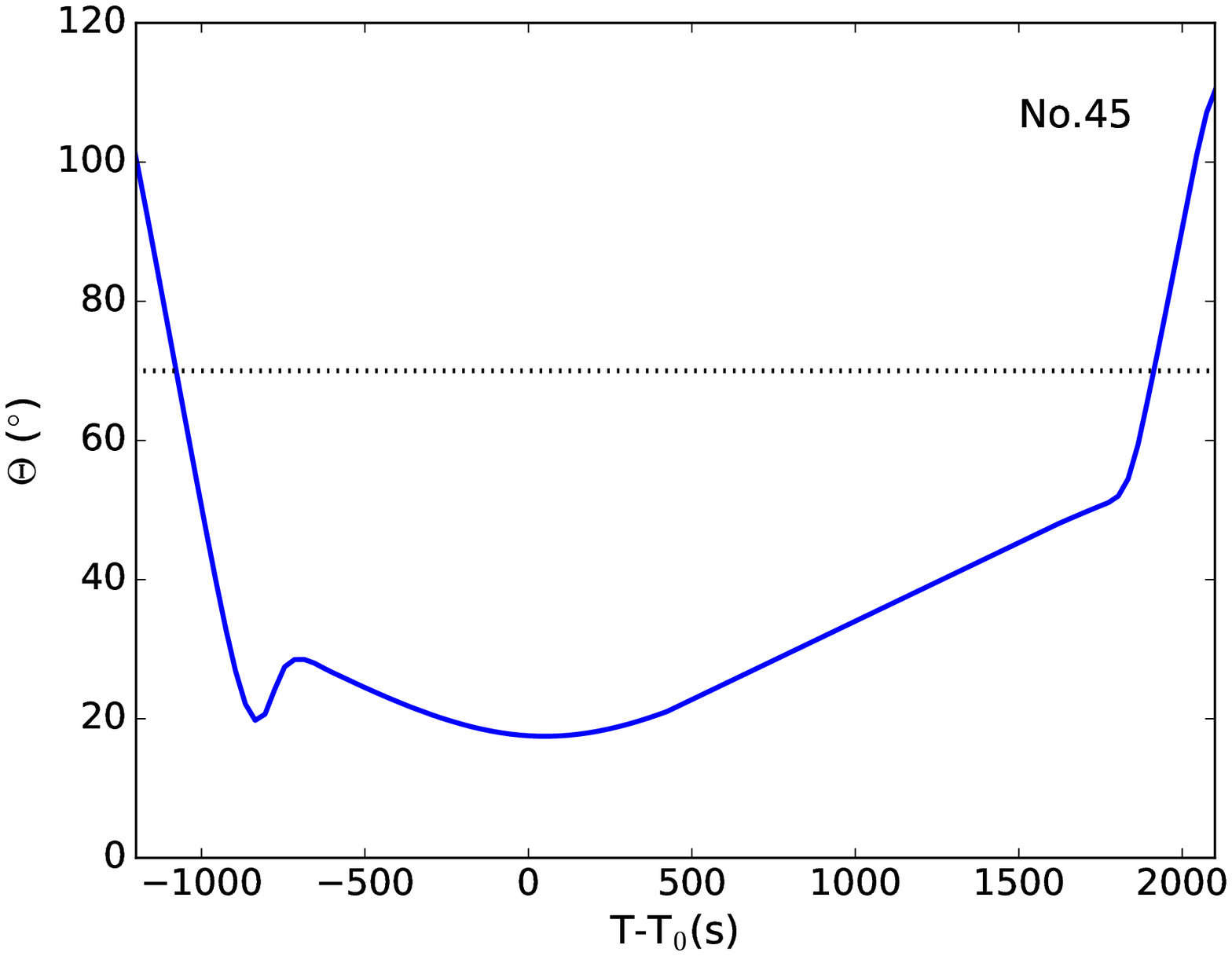}%
\includegraphics[angle=0,scale=0.3]{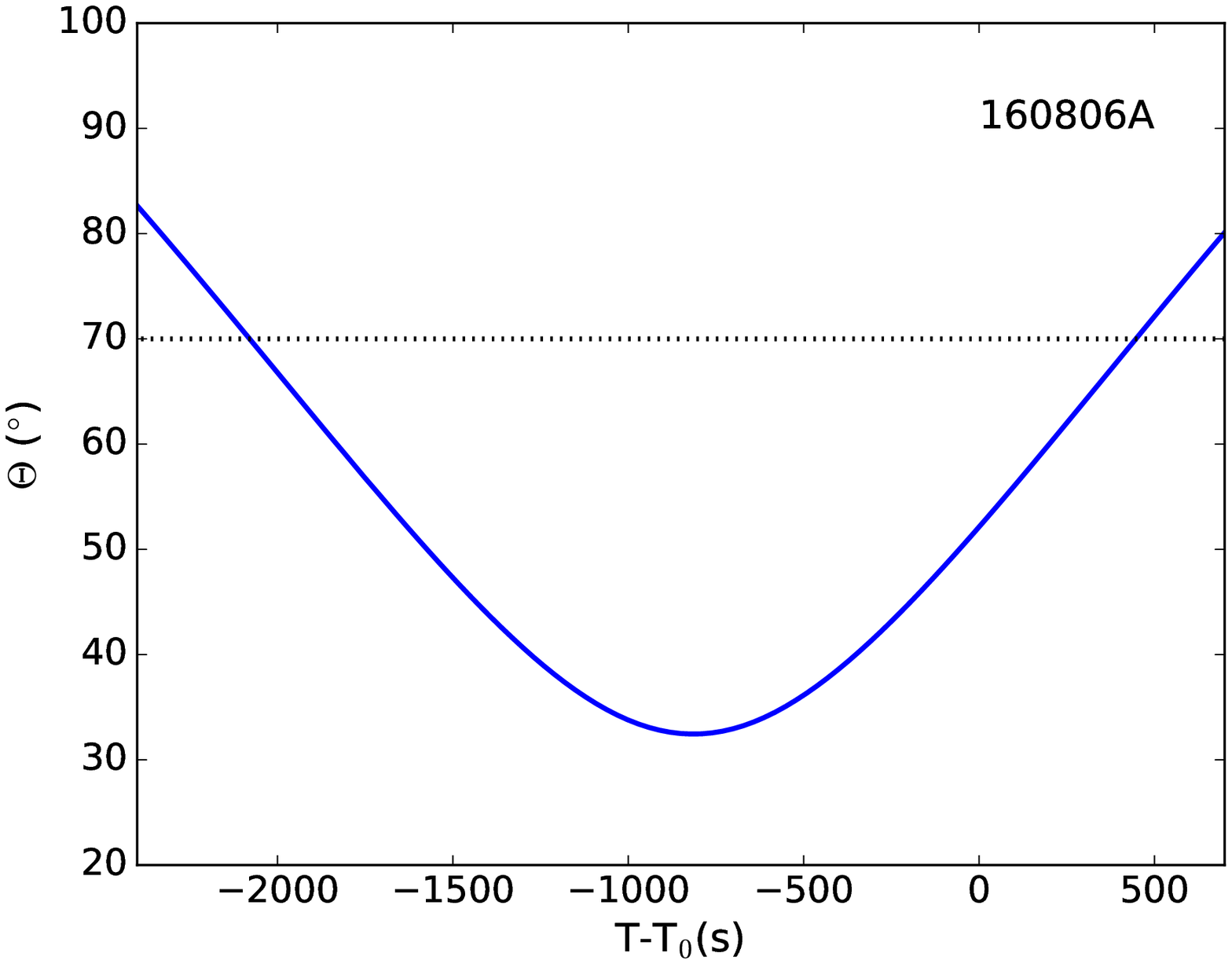}
\caption{The angular distance ($\Theta $) between the neutrino
position and \textsl{Fermi}-LAT boresight as a function of time
for the three IceCube track-like neutrino events. $T_0$ is the neutrino
detection time. The black horizontal dotted line represents $\Theta
= 70^{\circ}$.} \label{boresight}
\end{figure}

We note that three IceCube track-like neutrino events, e.g. number
23, 45 and 160806A,  locate at small angle ($< 70^{\circ}$) from
\textsl{Fermi}-LAT boresight at the neutrino detection time. In
other words, the region around the track-like neutrino events is
within the \textsl{Fermi}-LAT field of view during the  $\sim
1000$s before and after the neutrino detection. Therefore the
above three neutrinos are very suitable for searching for the
prompt GeV emission accompanying the neutrino emission. The
angular distance between neutrino position and \textsl{Fermi}-LAT
boresight (denoted as $\Theta$) versus time is shown in Figure
\ref{boresight}. The time intervals for \textsl{Fermi}-LAT data
analysis are selected with the criterion $\Theta < 70^{\circ}$,
which are also presented in Table \ref{lcondata}. The likelihood
analysis centered on each neutrino position result in upper limit
fluxes given in Table \ref{lcondata}. No new $\gamma $-ray point
sources are found within  $2R_{50}$ of the neutrinos position. For
the neutrino event number 23, there are four 3FGL sources within
$2R_{50}$, while for the other two neutrino events, there are no
sources within $2R_{50}$. The four 3FGL sources mentioned above
are very weak, and none of them shows any significant detection
over $\sim 1000$ s of \textsl{Fermi}-LAT observations. In brief,
we find no evidence of prompt GeV emissions following the IceCube
track-like neutrino events.

\section{Cross-correlation with \textsl{Swift} hard X-ray transient sources}
\begin{figure}
\centering
\includegraphics[scale=0.37]{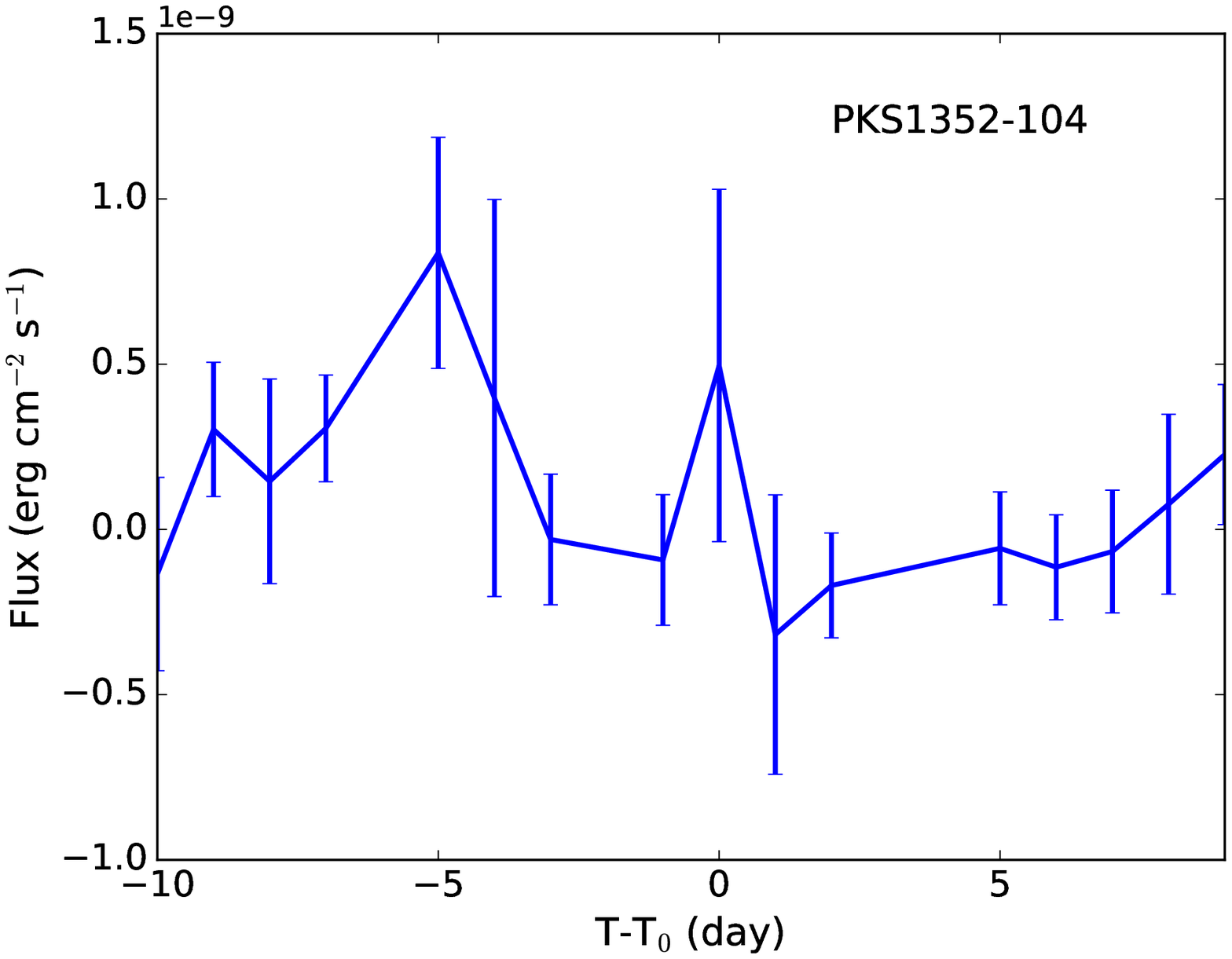}
\includegraphics[scale=0.37]{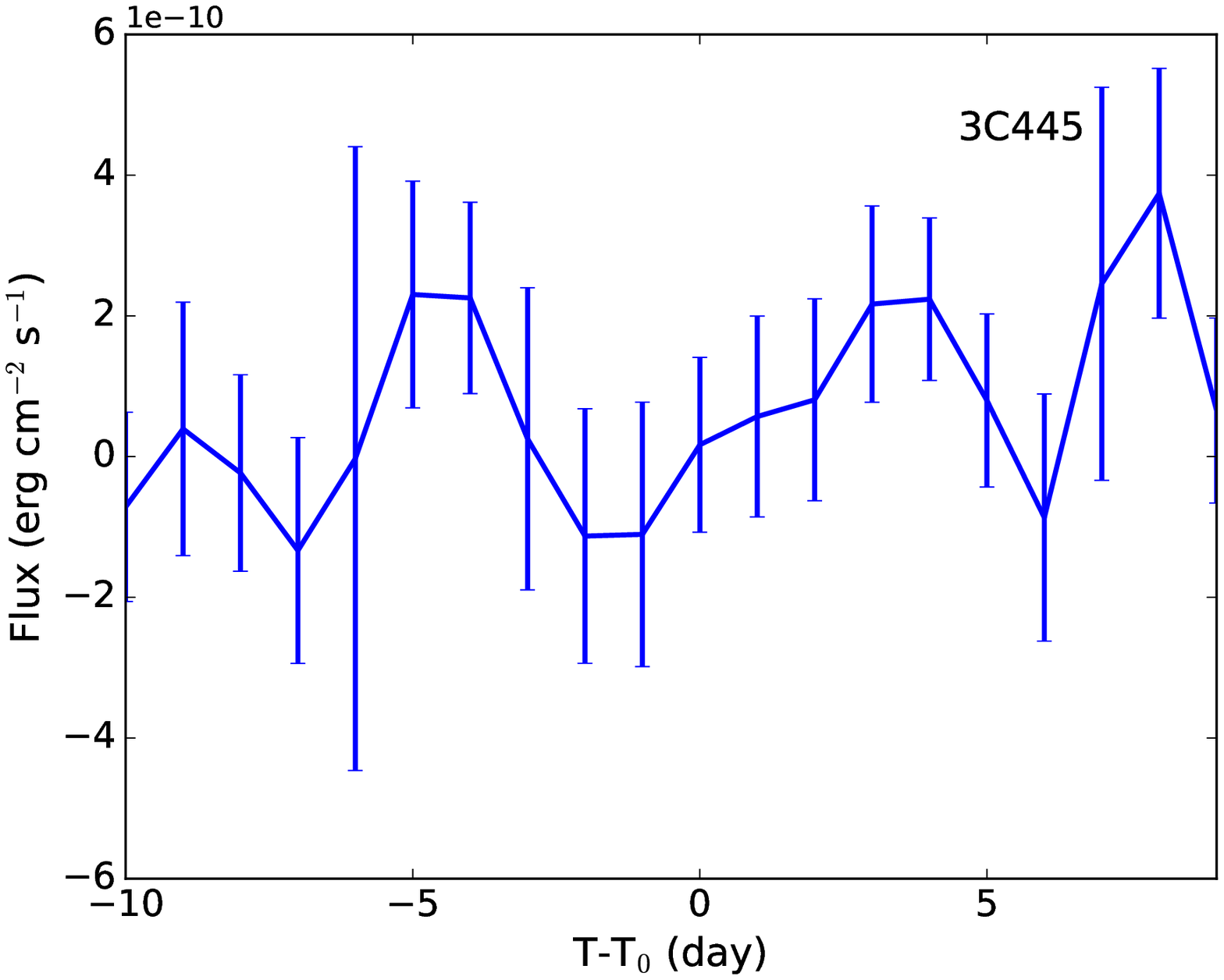}\\
\includegraphics[scale=0.37]{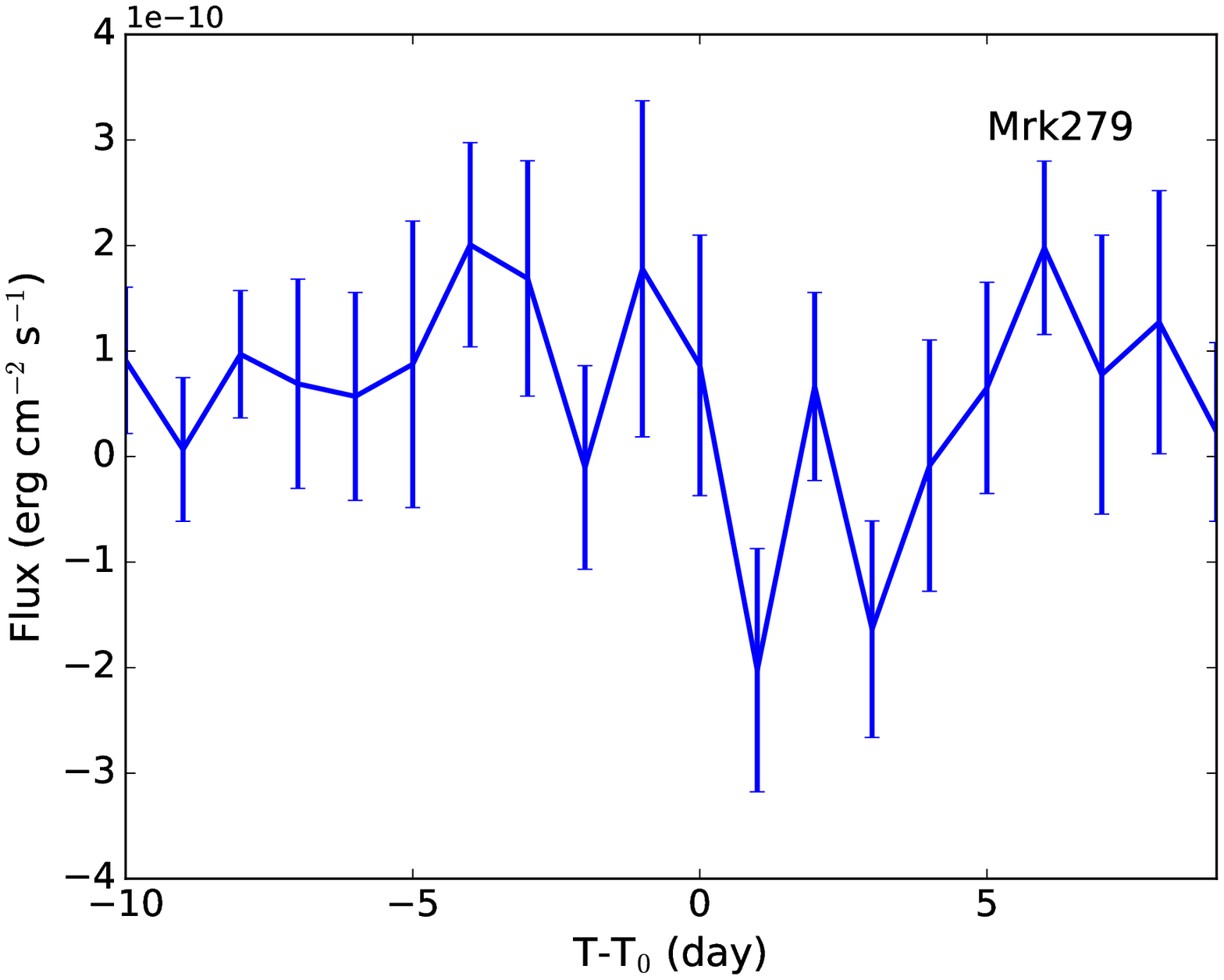}
\includegraphics[scale=0.37]{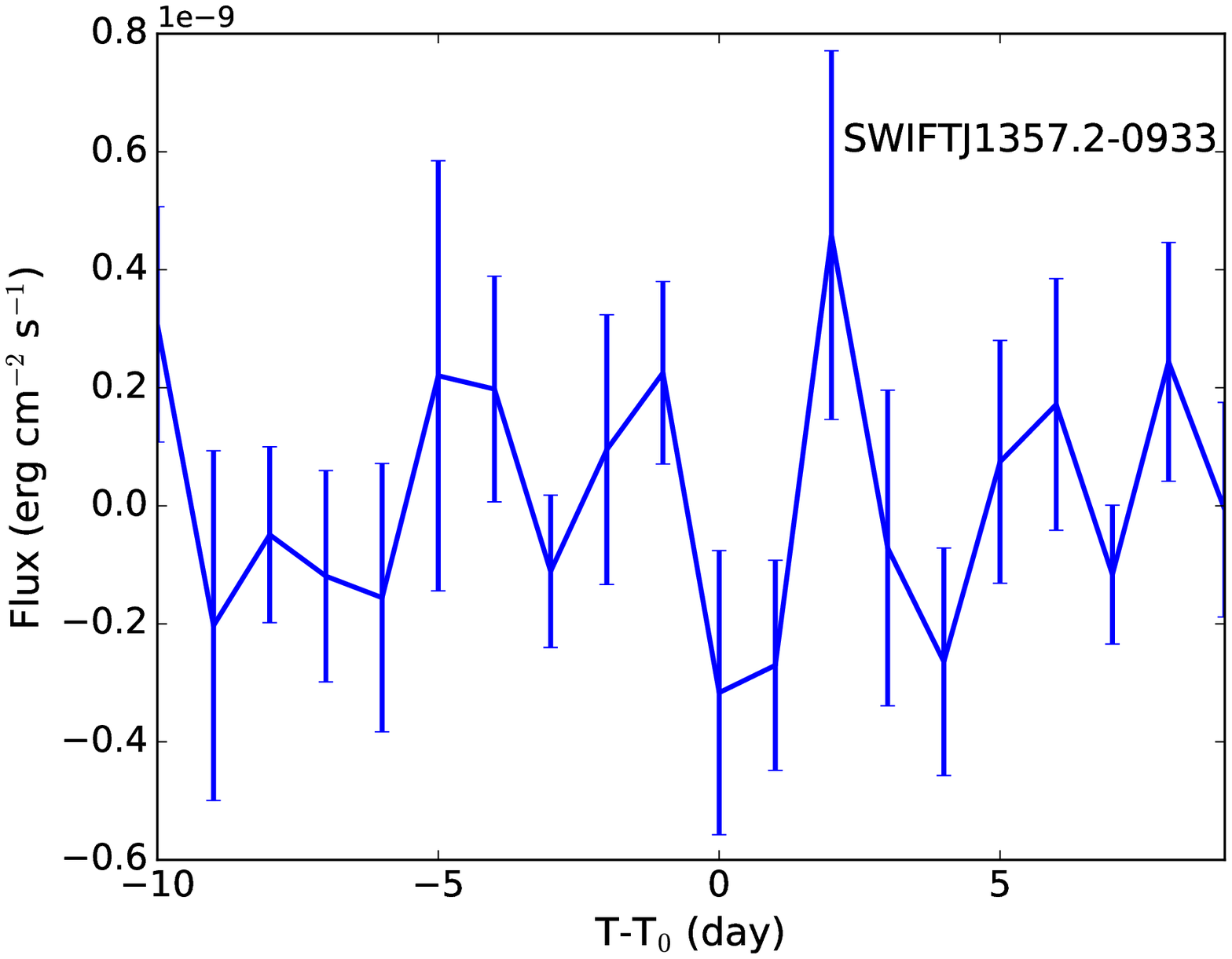}\\
\caption{The hard X-ray light curves of \textsl{Swift}/BAT
transient sources within $2R_{50}$ of the neutrino positions. The
neutrino detection time is denoted as $T_0$.} \label{BATtransient}
\end{figure}

The high-energy
photons from pion decay could be accompanied by X-ray emissions
that are produced by secondary electrons and positrons via, e.g.
synchrotron radiation in the magnetic fields of the source
\citep{2015arXiv151101530K,2016PhRvL.116g1101M}.
\textsl{Swift} Burst Alert Telescope (BAT) is very useful  for
discovering new X-ray transient sources or detecting the flux
variability of known X-ray sources
\citep{2005SSRv..120..143B,2013ApJS..209...14K}.  We thus make a
cross-correlation analysis between \textsl{Swift}/BAT transient
sources catalog and the IceCube track-like neutrino events. The
catalog includes  1009 X-ray transient sources (see
\url{http://swift.gsfc.nasa.gov/results/transients/}), including
Galactic and extragalactic sources.  None \textsl{Swift}/BAT
transient source is found inside the error box $R_{50}$ of the
IceCube track-like neutrino events. Within $2R_{50}$, four
\textsl{Swift}/BAT transient sources are in positional agreement
with the IceCube track-like neutrino events number 23, 44 and
47 (see Table \ref{BAT4sources}). To investigate their temporal
characteristics around the time that the corresponding neutrino
events are detected, we extract the day-bin light curves for these
X-ray sources. No significant flares are observed at the
neutrino detection time for the four X-ray transient sources, as
shown in Figure \ref{BATtransient}. Similarly, a chance
probability of $\sim 60\%$ for the positional coincidence is estimated using
Monte Carlo simulations  with the sample data randomized in right
ascension. Therefore, considering the insignificant spatial
coincidence and the observed temporal behavior of the X-ray
transients, we suggest that these \textsl{Swift}/BAT transient
sources are not in physical association with the IceCube
track-like neutrino events.

\section{Implications for constraining the neutrino source density }
The production of neutrinos are accompanied by high-energy
gamma-rays, { so the point source gamma-ray flux limits could, in
principle, provide useful constraints on the neutrino sources.}
For $pp$ collision mechanism of TeV-PeV neutrinos, one expect that
GeV gamma-ray flux lies at the power-law extrapolation of TeV
gamma-rays. The gamma-ray flux scales with the neutrino flux
through the relation $F_\gamma\simeq
2(E_\gamma/2E_\nu)^{2-p}F_\nu$ \citep{2013PhRvD..88l1301M},
assuming that the parent cosmic rays are produced with a power-law
spectrum, $dN_{CR}/dE_{CR}\propto E_{CR}^{-p}$. For $p\gamma$
mechanism, the flux of hadronic GeV gamma-rays depends on the
properties of soft target photons in the source. For simplicity,
below we assume that  $F_\nu (60-2000 {\rm TeV})=\eta F_\gamma
(0.2-100 {\rm GeV})$, with $\eta\simeq 0.5$ for $p=2$ in the $pp$
interaction model.  We assume that the sources are transparent to
gamma-rays, i.e. they are not the hidden sources in gamma-rays.
Our study has found that the sources of the track-like neutrino events
should be weak in $\gamma$-rays  in 0.2-100 GeV, even during the
neutrino emitting period.   Given a measured neutrino background
flux by IceCube, one can obtain a lower limit on the source number
density with the upper limit on the neutrino luminosity of
individual sources under the above assumptions. The observed
background neutrino flux implies a local energy production rate of
\begin{equation}
n_0 L_\nu \simeq 3\times 10^{44}{\rm \ erg \ Mpc^{-3} \ yr^{-1}}
\left(\frac{\xi_z}{3}\right)^{-1}\left(\frac{\sum \limits_i
E_{\nu,i}^2\Phi_{\nu,i}}{3\times 10^{-8}{\ \rm GeV \ cm^{-2} \
s^{-1} \ sr^{-1}}}\right),
\end{equation}
where $n_0$ is the local number density, $L_\nu$ is the averaged
neutrino luminosity of the each source throughout the universe,
$\xi_z$ is a dimensionless parameter that accounts for the
redshift evolution of the sources, and $\sum \limits_i
E_{\nu,i}^2\Phi_{\nu,i}$ is the all-flavor neutrino flux. The
upper limit gamma-ray flux for one year \textsl{Fermi}-LAT
observations is on average $ F_{\gamma} \la 7\times 10^{-13} $ erg
cm$^{-2}$ s$^{-1}$, so the limit on the neutrino flux of an
individual source is also $F_{\nu}\la 7\times 10^{-13}\eta$ erg
cm$^{-2}$ s$^{-1}$. As the neutrino source density is expected to
peak at $z\sim 1-2$ following the cosmic star formation rate, we
take the luminosity distance of these neutrino sources as
$d_L=10^{28}{\rm \ cm}$
\citep{2016ApJ...825..148C}\footnote{Though different objects show
different  evolution scenarios, hence different $\xi_z$, FSRQs and
star-forming galaxies both have a space density peaking at modest
redshift $z \approx  1$. Thus the luminosity distance of $D_L =
10^{28}$ cm is a plausible assumption in the ensuing discussion.}.
Then we obtain an upper limit of the neutrino luminosity of an
individual source
\begin{equation}
L_\nu\la 4\pi d_L^2 F_{\nu}\simeq 9\times 10^{44}\eta {\rm \ erg\
s^{-1}}.
\end{equation}
Thus, a lower limit on the continuous source density under the above assumptions may be written as
\begin{equation}
n_0\ga 10^{-8} {\rm \
Mpc^{-3}}\eta^{-1}\left(\frac{\xi_z}{3}\right)^{-1}\left(\frac{F_\gamma}{7\times
10^{-13} {\rm \ erg \ cm^{-2} \ s^{-1}}}\right)^{-1}.
\end{equation}
For short-term transient neutrino sources, the upper limit
gamma-ray flux for $\sim 1000$ s observations is about $
F_{\gamma} \approx 5 \times 10^{-10} $ erg cm$^{-2}$ s$^{-1}$, so
the energy released in neutrinos per event should be smaller than
$7 \times 10^{50}{\rm \eta}$ erg. Using a similar approach, we
find a lower limit on the event rate of the transients, i.e.,
\begin{equation}
\dot{n}_0\ga 4\times10^{2} {\rm \ Gpc^{-3}
yr^{-1}}\eta^{-1}\left(\frac{\xi_z}{3}\right)^{-1}\left(\frac{F_\gamma}{5\times
10^{-10} {\rm \ erg \ cm^{-2} \ s^{-1}}}\right)^{-1}.
\end{equation}
{We would like to stress that  the above lower limits  are
obtained based on the assumptions mentioned at the beginning of
this section. The validity of these assumptions depends heavily on
the relation between photon and neutrino fluxes.} The limits are
useful for constraining the source models. {FSRQs have a
number density of $\sim 10^{-9}{\ \rm Mpc^{-3}}$ and a faster
redshift evolution than the cosmic star formation rate
\citep{2012ApJ...751..108A,2014ApJ...780...73A}. Taking $\xi_z
\simeq 8.4$ for FSRQs, they are only marginally consistent with
the above constraint.} Starburst galaxies, one the other hand,
have a number density of $\sim 10^{-4}{\ \rm
Mpc^{-3}}$\citep{2012ApJ...755..164A}, so they fully satisfy the
above constraints for a large parameter space of $\eta$. For
short-term transient sources, since high-luminosity GRBs have a
density of $\sim 1 \ {\rm Gpc^{-3}\ yr^{-1}}$, one can rule out
GRBs as the main contributing sources of these neutrinos if
$\eta<100$ (i.e. the neutrino flux at TeV-PeV energies is a factor
of  $<100$ larger than that in GeV gamma-rays). Since
low-luminosity GRBs have a density of $200-1000 {\ \rm Gpc^{-3}\
yr^{-1}}$, they can not be ruled out by our \textsl{Fermi}-LAT
data analysis. {We note that the constraints on the source
density are generally consistent with the results obtained by
using the non-detection of high-energy neutrino multiplets in the
IceCube data \citep{2014PhRvD..90d3005A,2016arXiv160701601M}.}

\section{Conclusions and Discussions}
By using \textsl{Fermi}-LAT  observations, we  searched for
$\gamma$-ray transient emission on the timescales of hours to months
coincident with the IceCube track-like neutrino events above 60 TeV.
The null result suggests that any associated gamma-ray flares must
be at least one order of magnitude dimmer than that of the blazar
PKS B1424-418, for which a PeV cascade-like neutrinos is claimed to be
associated at 95\% confidence level. For three track-like neutrinos that occurred within
the field of view of \textsl{Fermi}-LAT at the time of the
neutrino detection, we also searched for  prompt GeV emission
coincident in time with these neutrinos. No significant GeV
emissions associated with these  neutrino events are found. A few
3FGL $\gamma $-ray objects locate within $2R_{50}$ of the neutrino
position, but the probability for chance coincidence is large.
They are also too weak in gamma-ray emission to be reconciled with
the neutrino emission. Based on the  non-detections of GeV
emissions and some assumptions (see Section 4), the inferred local number density for continuous
emitting sources to produce high-energy neutrinos should be $ n_0
\ga 10^{-8}$ Mpc$^{-3}$ by assuming a flat gamma-ray spectrum
resulted from the $pp$ mechanism for neutrinos. Similarly, for
transient sources, we obtain an event rate of $\dot{n}_0\ga
4\times10^{2} {\rm \ Gpc^{-3} yr^{-1}}$. We also searched for
possible hard X-ray transients observed by \textsl{Swift}/BAT that
are coincident with the track-like neutrino events, but no X-ray
flares are found to be spatially and temporally coincident with
these neutrino events.

Some alternative explanations for non-detection of 0.2-100 GeV
emission accompanying the IceCube track-like neutrino events are
possible. For example, if the high-energy photons produced in the
pion mesons process can not escape freely from the source region
(i.e. the hidden sources in gamma-rays), they would not suffer
from the above constraints. Another possibility is that the
gamma-ray luminosity at GeV energies is far below that of the
TeV-PeV neutrinos in the $p\gamma$ scenario when the energy
threshold of cosmic rays for pion production in interactions with
radiation fields is too high. Future prompt follow-up observations
in TeV energies by  Imaging  Cherenkov Telescopes, such as CTA,
HAWC and LHAASO,  would be useful to test the latter possibility.

\section*{Acknowledgments}
We thank Ruo-Yu Liu and Xiao-Chuan Chang for useful discussions.
We also acknowledge a  constructive report from the referee. This
work has made use of data and software provided by the
\textsl{Fermi} Science Support Center, and \textsl{Swift}/BAT
transient monitor results provided by the \textsl{Swift}/BAT team.
This work is supported by the 973 program under grant
2014CB845800, the NSFC under grants 11625312 and 11273016.

\clearpage

\begin{table}
\centering
\begin{threeparttable}
\caption{Upper limit gamma-ray fluxes of the track-like neutrino
events as observed by \textsl{Fermi}-LAT on the timescales of 12 hours
and one year. The first column is the neutrino ID. The second
column represents the energy of each neutrino event. The third and
forth columns describe the positions of the neutrinos. The fifth
column represents the median angular error $R_{50}$. The last two
columns are the upper limit fluxes (0.2-100 GeV) over 12 hours and
one year observations by \textsl{Fermi}-LAT around the neutrino
detection time, respectively. The measured fluxes of the
gamma-ray flare of PKS B1424-418 on the timescales of 12 hours and one
year are also shown for comparison.}
\begin{tabular}{lcccccc}
\hline
\hline
ID& Energy    &R.A.   &Dec.& Angular error & Flux($\times 10^{-8}$) &    Flux($\times 10^{-10}$)  \\
&   (TeV)&  ($^{\circ}$)& ($^{\circ}$) & ($^{\circ}$)   &   ph cm$^{-2}$ s$^{-1}$ &ph cm$^{-2}$ s$^{-1}$ \\
\hline
3   &$78.7^{+10.8}_{-8.7}$  &127.9  &   -31.2   &   1.4       &   8.05   &   3.75  \\
5   &   $71.4^{+9.0}_{-9.0}$    &110.6  &   -0.4    &   1.2   &   5.94    &   17.4\\
13  &   $253^{+26}_{-22}$   &67.9   &   40.3    &   1.2   &   9.25    &   8.17\\
23  &   $82.2^{+8.6}_{-8.4}$    &208.7  &   -13.2   &   1.9   &   9.80 &   19.9     \\
38  &   $200.5^{+16.4}_{-16.4}$ &93.34  &   13.98   &   1.2   &   18.1   & 35.0    \\
44  &   $84.6^{+7.4}_{-7.9}$    &336.71 &   0.04    &   1.2   &   32.4    & 4.19   \\
45  &   $429.9^{+57.4}_{-49.1}$ &218.96 &   -86.25  &   1.2   &   5.71    & 14.5    \\
47  &   $74.3^{+8.3}_{-7.2}$    &209.36 &   67.38   &   1.2   &   3.81    & 7.40    \\
55\tablenotemark{a}  &   $2600\pm 300$&110.34    &   11.48   &   0.27    &   11.3   &  8.61    \\
160427A\tablenotemark{c} & ..  &240.57 &   9.34    &   0.6\tablenotemark{b}   &   9.95   &   20.8     \\
160731A\tablenotemark{c} & ..  &215.109    &   -0.4581 &   0.35\tablenotemark{b}  &  12.5   &   24.7   \\
160806A\tablenotemark{c} & ..  &122.81 &   -0.8061  &   0.5\tablenotemark{b}  &   3.22    &   5.15\\
\hline
PKS B1424-418   &       &   &   &   &        $35.4\pm9.92$ &   $5889\pm60$\\
\hline
\end{tabular}
\textbf{Notes.}
\begin{tablenotes}
\item[a] Neutrino event number 55 is a PeV event with
$R_{50}\approx 0.27^{\circ}$ \citep{2015ATel.7856....1S}. \item[b]
160427A--\citet{2016GCN..19363...1B};
160731A--\url{http://gcn.gsfc.nasa.gov/notices_amon/6888376_128290.amon};
160806A--\citet{2016GCN..19787...1C}. \item[c] For the neutrino
events with number 160427A, 160731A, and 160806A, their deposited energy are
not given in the literatures or GCN Circulars.
\end{tablenotes}
\label{lcdata}
\end{threeparttable}
\end{table}

\begin{table}
\centering \caption{3FGL sources that are found around the
positions of the IceCube track-like neutrino events. The energy fluxes
(in unit of $10^{-11}$ erg cm$^{-2}$ s$^{-1}$) for one year
\textsl{Fermi}-LAT  observations are presented for sources only
with $TS > 25$  (under each 3FGL source name respectively).}
\begin{tabular}{lccccc}
\hline
\hline
ID  &   $R_{50}$    &   $2R_{50}$   &       &       &       \\
\hline
3   &       &   J0825.8-3217    &       &       &       \\
\hline
5   &   J0725.8-0054    &   J0721.5-0221    &   &   &       \\
    &   $1.09\pm0.19 $  &   &       &   &       \\
\hline
13  &       &   J0423.8+4150    &       &       &       \\
    &       &   $2.82\pm0.31 $  &       &       &       \\
\hline
23  &&J1349.6-1133&J1351.8-1524&J1355.0-1044&J1400.5-1437   \\
    &       &   $4.61\pm0.23 $  &   &&$0.77\pm0.17 $    \\
\hline
44  &   J2227.8+0040    &   J2223.3+0103    &   &   &   \\
    &   $0.62\pm0.16 $  &   &       &       &       \\
\hline
47  &       &   J1404.8+6554&   &       &\\
    &       &   $0.32\pm0.09 $ &    &   &   \\

\hline
\end{tabular}
\label{FGLdata}
\end{table}

\clearpage

\begin{table}
\centering \caption{Upper limit gamma-ray fluxes of the three
IceCube track-like neutrino events that locate within the
\textsl{Fermi}-LAT's field of view at the neutrinos detection
time. The first column is the neutrino ID, the second column is
the time interval when the angular distance between the neutrino
position and the \textsl{Fermi}-LAT boresight is less than
$70^{\circ}$, and the last column is the upper limit flux in
0.2-100 GeV.}
\begin{tabular}{lcccc}
\hline
\hline
ID  & Time$+T_0$(s) &   Flux    \\
    & & $10^{-7}$ ph cm$^{-2}$ s$^{-1}$     \\
\hline
23  &[-660,950] &   4.39\\
45  &[-1070,1910]   &   2.26    \\
160806A &[-2080,440]    &   5.56 \\

\hline
\end{tabular}
\label{lcondata}
\end{table}

\clearpage

\begin{table}
\centering \caption{\textsl{Swift}/BAT transient sources around
$2R_{50}$ of the IceCube track-like neutrino events. The last
column is the angular separation between the positions of the
neutrino and the X-ray sources. 'LMXB' means low-mass X-ray
binary.}
\begin{tabular}{lcccc}
\hline
\hline
ID  & Angular error($^{\circ}$) &   X-ray sources & type &separation($^{\circ}$) \\
\hline
23  &1.9    &   PKS 1352-104 & Blazar &2.52 \\
    &       &   Swift J1117.1-0933 & LMXB & 3.71\\
44  &1.2    &   3C 445 & Seyfert Galaxy &2.27   \\
47  &1.2    &   Mrk 279 & Seyfert Galaxy  &1.98 \\

\hline
\end{tabular}
\label{BAT4sources}
\end{table}

\end{document}